\documentclass[aps, twocolumn, showpacs, amsmath]{revtex4}
\usepackage{dcolumn}% Align table columns on decimal point
\usepackage{bm}% bold math
\usepackage{graphicx}% Include figure files
\usepackage{color}
\usepackage{subfigure}
\usepackage{hyperref}

\DeclareGraphicsExtensions{.jpg,.pdf, .mps, .png, .eps, .ps, .EPS,.gif}

\begin{document}
\def\be{\begin{equation}}
\def\ee{\end{equation}}

\def\bc{\begin{center}} 
\def\ec{\end{center}}
\def\bea{\begin{eqnarray}}
\def\eea{\end{eqnarray}}
\newcommand{\avg}[1]{\langle{#1}\rangle}
\newcommand{\Avg}[1]{\left\langle{#1}\right\rangle}

\def\ie{\textit{i.e.}}

\title{Multiplex networks with heterogeneous activities of the nodes }

\author{Davide Cellai$^{1,2}$  and  Ginestra Bianconi$^3$}

\affiliation{
$^1$ Idiro Analytics, Clarendon House, 39 Clarendon Street, Dublin 2, Ireland\\
$^2$ MACSI, Department of Mathematics and Statistics, University of Limerick, Ireland\\
$^3$ School of Mathematical Sciences, Queen Mary University of London, London, E1 4NS, United Kingdom}
\begin{abstract}

In multiplex networks with a large number of layers, the nodes can have different activities,  indicating the total number of layers in which the nodes are present. Here we  model multiplex networks with heterogeneous activity of the nodes and we study their robustness properties.  We introduce a percolation model where nodes need to belong to the giant component only on the layers where they are active ({\ie} their degree on that layer is larger than zero).  We show that when there are enough nodes active only in one layer, the multiplex becomes more resilient and the transition becomes continuous. We find that multiplex networks with  a power-law distribution of node activities  are more fragile if the distribution of activity is broader. 
We also show that while positive correlations between node activity and degree can enhance the robustness of the system, the phase transition may become discontinuous, making the system highly unpredictable.

\end{abstract}
%\pacs{64.60.aq, 64.60.Cn, 89.75.Hc} 
%\pacs{89. 75-k, 89. 75. Fb, 89. 75. Hc}

\maketitle
\section{Introduction}
 Multilayer networks \cite{PhysReports,Kivela,Goh_review} {are} formed by several interacting networks.
{They } describe a large variety of complex systems. Examples are found in social \cite{Thurner}, technological, communication, transportation systems \cite{Mucha,Boccaletti,Vito} but also  in biological networks of the cell, or in the brain \cite{Bullmore2009,Makse,Liaisons}.
In the last fifteen years a lot of attention has been devoted to understand the interplay between structure and dynamics in single networks \cite{crit,Dynamics}. {In recent years, it has become} clear that most of the networks are not isolated and that for understanding the function of complex systems as different as complex infrastructures or the brain it is important to investigate the role of their multilayer structures \cite{PhysReports,Kivela}.
In particular, ample debate has been devoted to characterize the robustness of multilayer networks \cite{Havlin1,Havlin2,Gao1,Gao2,Gao3,Son,Dorogovtsev,Goh,Kcore,JSTAT,cellai2013,Cellai,Kabashima,BD1,BD2,Makse,Boguna,Kahng}.
It has been shown that, in the presence of interdependencies, the robustness of a multilayer network can be significantly affected.
{Multilayer networks can be much more vulnerable to random damage with respect to considering only their  single layers taken in isolation \cite{Havlin1,Son,Dorogovtsev}.

In the presence of interdependencies, the notion of mutually connected component has been introduced, meaning that each pair of nodes in the mutually connected component must be connected by a path on each and every layer, internal to this component.
This definition is motivated by the fact that in these interdependent systems a node is not functional if any of its interdependent nodes in the other layers is not  functional.
Therefore, the largest (giant) mutually connected component (MCGC) describes the robustness of the system.}
This component  has   a discontinuous phase transition as a function of the initial random damage inflicted to the nodes of the network, and   close to this transition the system is affected by dramatic cascades of failure events.
This transition can be studied on multilayer networks of different nature, including multiplex networks \cite{Havlin1,Havlin2,Son,Dorogovtsev} and network of networks \cite{Gao1,Gao2,Gao3,BD1,BD2}.
Network of networks are formed by different interacting networks, as the molecular networks in the cell, in which every node is a different type of biological molecule.
Multiplex networks \cite{Boccaletti,Battiston,Vito,PRE,Growth1,Math,Menichetti,Garlaschelli}, instead,  are multilayer networks in which the same set of nodes interact through different networks. Example of multiplex networks are social networks, in which individual are connected by different types of social ties, transportation networks, like airport networks in which each airport is connected to other airports though connections of different airline companies, or collaboration networks in which scientists collaborate on different topics and eventually cite each other. Several works have focused on modeling multilayer networks \cite{PRE,Growth1,Math,Menichetti,Garlaschelli}. In particular a very useful approach employs statistical ensembles  which are able to generate a large variety of multiplex network topologies with desired level of structural correlation \cite{PRE,Menichetti,Garlaschelli}.
In networks of networks as well as in multiplex networks the possible correlations existent in this structure can strongly affect their robustness properties \cite{PhysReports,Goh,Liaisons, Makse}.

An interesting result has recently shown that multiplex networks are characterized by the fact that not all the nodes are connected in every layer. In fact, many networks have been shown \cite{Vito} to have heterogeneous   activity of the nodes. The activity of a node is given by the   number of layers in which the node is at least connected to another node.  The activity of the nodes has been found to be  broadly distributed in a variety of multiplex network data sets \cite{Vito}.
Moreover, the activity of the nodes has been seen to correlate in average with the degree of single node in given layers, meaning that on average nodes that are present in many layers have also typically high degree in the single layers in which they are active \cite{Vito}.
{The heterogeneous distribution of activities of the nodes and its relation with the mean degree in single layers is frequently observed in real-world multiplex networks and may encode relevant information \cite{Vito}.
Besides, multiplex networks with heterogeneous activities can be useful theoretical tools to address other problems like modular structures in single networks \cite{colomer2014}.

Here we characterize multiplex networks ensembles with heterogeneous activities of the nodes and we investigate their robustness properties, assuming mutual interdependencies among each layer.}
This ensemble of multiplex networks with heterogeneous activity of the nodes can be used to model realistic multiplex network structures, on top of which dynamical processes may occur.
Moreover, ensemble of multiplex networks with heterogeneous activities of the nodes can be used to estimate the role that correlation have on their robustness properties.
Here we  find that heterogeneous activity of the nodes can decrease  the robustness of networks in the presence of interdependencies. Nevertheless the correlation between the activity of the nodes and their degree within single layers has the opposite effect and can improve the robustness of multiplex networks.

\section{The multiplex network ensemble with heterogeneous activity of the nodes}
\label{ensemble}
\begin{figure}
	\includegraphics[width=0.99\columnwidth]{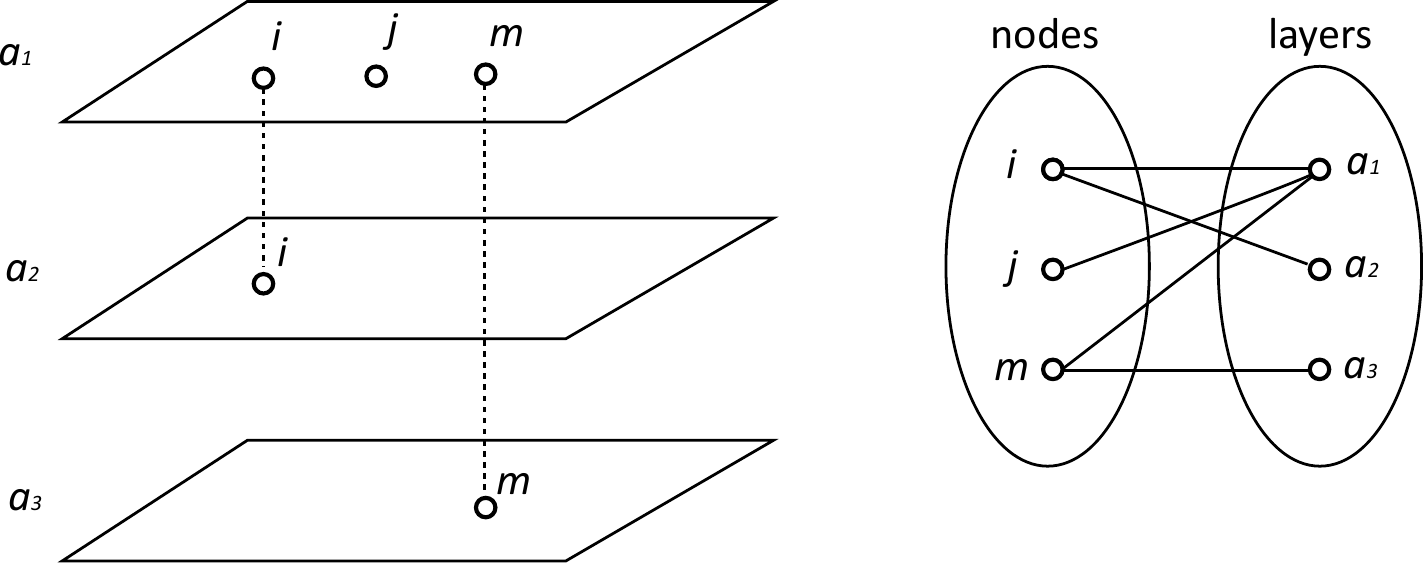}
	\caption{{Sketch of a multilayer network with node activity: nodes are not active in every layer (left), so the system can be represented by a bipartite graph associating each node to the layers where it appears (right).}}
	\label{fig:bipartite-sketch}
\end{figure}

A multiplex network of $N$ nodes $i=1,2,\dots, N$ and $M$ layers $\alpha=1,2,\ldots,M$ is completely specified when the $M$ adjacency matrices of elements $a_{ij}^{\alpha}=1$ if node $i$ is linked to node $j$ in layer $\alpha$, and otherwise $a_{ij}^{\alpha}=0$ are given. Every node of the multiplex network is labelled as $(i,\alpha)$, indicating that is the $i$-th node in layer $\alpha$. The replica nodes of the node $(i,\alpha)$ are defined as all the nodes labelled as $(i,\alpha')$ in layers $\alpha'\neq \alpha$ \cite{BD1}.   
Interestingly, it has been observed from data \cite{Vito}, that in many networks not all the nodes are active ({\ie}  are connected to at least another node)  in each layer.
Let us define the activity $B_i$ of a node $i$ as the number of layers where node $i$ has non-zero degree.
The activities of the nodes are broadly distributed \cite{Vito} and they  can be fitted by a power-law $P(B)\simeq B^{-\delta}$ with $\delta\in [1.5,3.0]$.
This implies that for some multiplex networks the bipartite network between nodes and layers described by the activity adjacency matrix can be either dense $\delta\leq 2$ or sparse $\delta>2$ but the typical number of layers in which a node is active is always subject to unbound fluctuations.
In order to characterize fully the activities of the nodes in each layer, from the $M$ adjacency matrices ${\bf a}^{\alpha}$  it is possible to construct a $N\times M$ activity matrix  ${\bf b}$ of elements $b_{i,\alpha}$ (Fig.~\ref{fig:bipartite-sketch}). This matrix can be viewed as an adjacency matrix between nodes and layers indicating if node $i$ is active in layer $\alpha$ ($b_{i,\alpha}=1)$ or not ($b_{i,\alpha}=0$).
%
%For an undirected multiplex network the node $i$ is active in layer $\alpha$  if it  has a positive degree in layer $\alpha$, i.e. $k_i^{\alpha}>0$.
%Therefore  we have  
%\bea
%b_{i,\alpha}=1-\delta_{0,k_i^{\alpha}}=1-\delta_{0,\sum_{j=1}^N a_{i,j}^{\alpha} },
%\eea
%where $\delta_{x,y}$ indicates the Kronecker delta.
The activity $B_i$ of a node $i$ can be therefore expressed in terms of the matrix ${\bf b}$ as in the following,
\bea
B_i=\sum_{\alpha=1}^Mb_{i,\alpha}.
\eea
The layer activity $N_{\alpha} $ has been defined in \cite{Vito,PhysReports} and  is given by the number of nodes active in layer $\alpha$, i.e.
\bea
N_{\alpha}=\sum_{i=1}^Nb_{i,\alpha}.
\eea
It is therefore possible to construct ensemble of multiplex networks with heterogeneous activity of the nodes in the subsequent manner: first we construct the network between layers and nodes described by the adjacency matrix $b_{i,\alpha}$ indicating if node $i$ is active in layer $\alpha$, then we can construct in each layer a network between the active nodes of the layer with a degree distribution $P^{\alpha}(k)$.
In the case in which   $P^{\alpha}(k=0)=0$,  we have that a node is active in a  given layer if it is connected at least to another node in the same layer, i.e.
\bea
b_{i,\alpha}=1-\delta_{0,k_i^{\alpha}}=1-\delta_{0,\sum_{j=1}^N a_{i,j}^{\alpha} },
\eea
where $\delta_{x,y}$ indicates the Kronecker delta.

In order to construct the activity matrix $b_{i,\alpha}$ we can consider a microcanonical ensemble in which  the  activities of the nodes $B_i$ and the  layer activities $N_{\alpha}$ are fixed.
This ensemble, called also the configuration model of a bipartite network, can be constructed by 
 maximizing the entropy $S$ of the activity networks  given by 
 \bea
 S=-\sum_{{\bf b}}P({\bf b})\log P({\bf b}) 
 \eea
 where $P({\bf b})$ is the probability of a given activity network in the ensemble under the constraints that the  node activity and layer activity are kept constant.
In this ensemble the probability of a matrix ${\bf b}$ is given by 
\bea
P({\bf b})=\frac{1}{Z}\prod_{i=1}^N\delta\left(\sum_{\alpha=1}^Mb_{i,\alpha},B_i\right)\prod_{\alpha=1}^M\delta\left(\sum_{i=1}^Nb_{i,\alpha},N_{\alpha}\right),
\label{Pben}
\eea
where $Z$ is the normalization constant also called the partition function of the ensemble.
In this  ensemble  the probability $p_{i,\alpha}$ that a node $i$ is active in layer $\alpha$ is expressed in terms of the Lagrangian multipliers $\lambda_i$, $\omega_{\alpha}$, \ie
\bea
	p_{i\alpha}=\frac{e^{-\lambda_i-\omega_{\alpha}}}{1+e^{-\lambda_{i}-\omega_{\alpha}}}.
	\label{p-lagrangian-mult}
\eea 
The Lagrangian multipliers are fixed by the conditions 
\bea
\sum_{i}p_{i,\alpha}&=&N_{\alpha}\nonumber \\
\sum_{\alpha}p_{i,\alpha}&=&B_i
\eea
Finally the entropy \cite{GCP,Anand2010,Anand2009} $S$ of the ensemble of activity networks is given by 
\bea
S={\cal S}-\Omega
\eea
where ${\cal S}$ and $\Omega$ are given by 
\bea
{\cal S}&=&-\sum_{i\alpha}\left[p_{i,\alpha}\ln p_{i,\alpha}+(1-p_{i,\alpha})\ln(1-p_{i,\alpha})\right],\nonumber \\
\Omega&=&-\sum_i \ln\left[\frac{1}{B_i!}B_i^{B_i}e^{-B_i}\right]-\sum_{\alpha}\ln\left[\frac{1}{N_{\alpha}!}N_{\alpha}^{N_{\alpha}}e^{-N_{\alpha}}\right].\nonumber 
\eea
We assume here that  bipartite networks between nodes and layers, characterized by the adjacency matrix $b_{i,\alpha}$ is uncorrelated, i.e. we assume that
 the activities $B_i$ of the nodes are not correlated with the sizes of active nodes  $N_{\alpha}$ on the layers where the nodes are active.
 In this hypothesis the probabilities $p_{i,\alpha}$ are proportional to the product of $B_i$ and $N_{\alpha}$ which are the degrees of the mentioned bipartite network, and the following condition on the  the maximal activity of the nodes $B_{max}$ and the maximal layer activity $N_{max}$ must be satisfied, {\ie}
\bea
	\frac{B_{max}N_{max}}{\sum_i B_i}\ll 1.
	\label{sparse}
\eea
This condition is the condition that in a bipartite network corresponds to the one imposing a structural cutoff on single uncorrelated networks (networks without degree-degree correlations) \cite{cutoff}.
In this case we have the simple expression
\bea
	{p_{i,\alpha}}=\frac{B_iN_{\alpha}}{\sum_{\beta}N_{\beta}},
	\label{p-factorizable}
\eea
with $\sum_{\beta}N_{\beta}=\sum_i B_i=\avg{N}M=\avg{B}N$.

As we said, this argument  regards lack of correlation between activities and number of active nodes on the layer.
In Sec.~IV, we will examine a different type of correlation, namely the one between activities and the degree distribution on the layer.

Once the activity network is constructed, in order to construct the multiplex network, we assign to each  active node of layer $\alpha$ a degree within the layer.
If the activity of the nodes are uncorrelated with the degree in each layer the degrees $k_{i,\alpha}$ of the nodes $i$ in layer $\alpha$ are drawn form  the degree distribution $P^{\alpha}(k)$ and the networks  of each layer $\alpha$ are generated by the configuration model with degree distribution $P^{\alpha}(k)$.
In other words, the probability of the multiplex network degree sequences ${\bf k}=\{k_{i,\alpha}\}_{i=1,2\ldots, N;\alpha{=1,2},\ldots, M}$ is given by
\bea
	P(\{ {\bf k}\}|{\bf b})=\prod_{i,\alpha}\left[P^{\alpha}(k_i^{\alpha}){b_{i,\alpha}}+\delta(k_i^{\alpha},0)(1-b_{i,\alpha})\right].
\label{Pkb}
\eea
Instead, if there is a correlation between the degree of the nodes within a layer and the activity of the nodes, then we need to draw the degree of the nodes in each layer from a probability $P_{B_i}(k_i)$ which is a function of their activity $B_i$.

\section{Mutually connected component in a multiplex network with given distribution of activities of the nodes}

We consider the mutually connected giant component (MCGC) in a multiplex network with given distribution of activities of the nodes, described by the ensemble of multiplex networks introduced in section \ref{ensemble}.

The layers are interdependent, meaning that each node active in a given layer $\alpha$ is interdependent on its replica nodes in all the layers $\beta$ where the node is active.
In particular we will assume that a node $(i,\alpha)$ active in layer $\alpha$ belongs to the mutually connected giant component if 
\begin{itemize}
 \item[(a)] at least one neighbor $(j,\alpha)$ of node $(i,\alpha)$  belongs to the mutually connected giant component;
 \item[(b)] {  in each layer $\beta\neq\alpha$ } where the node $i$ is active, at least one neighbor  $(j,\beta)$ belongs to the mutually connected giant component.
\end{itemize}
On a locally tree-like multiplex network, this can be easily encoded in a message passing algorithm determining if a node $(i,\alpha)$ belongs to the mutually connected giant component \cite{PhysReports,Kabashima,Son,Mezard}.
The indicator function $S_{i\alpha}$ indicates if a nodes $(i,\alpha)$ belongs ($S_{i\alpha}=1$) or not ($S_{i\alpha}=0$) to the mutually connected giant component.
This indicator is determined by a set of ``messages" $\sigma^{\alpha}_{i\to j} $ that each node $(i,\alpha)$ active in a layer $\alpha$ sends to the neighboring nodes $(j,\alpha)$ in the same layer. Each message    $\sigma^{\alpha}_{i\to j} $ indicates if the node $(i,\alpha)$ belongs ($\sigma^{\alpha}_{i\to j}=1$) or not ($\sigma^{\alpha}_{i\to j}=0$) to the mutually connected component when the link to the node $(j,\alpha)$ is removed.

Here our goal is to characterize the size of the mutually connected giant component as a function of the probability $1-p$ that random nodes are damaged in the network. 
In order to characterize the damage initially inflicted to the network we indicate with $s_{i}=0,1$ if a node has been damaged ($s_{i}=0$) or not ($s_i=1$) in the multiplex network. 
Given the definition of the mutually connected component, the  message  $\sigma^{\alpha}_{i\to j} $ from a  node $(i,\alpha)$ to a node  $(j,\alpha)$ both active in layer $\alpha $ is therefore given by 
\bea
	\sigma_{i\to j}^{\alpha}&=&s_{i} \left[1-\prod_{\ell'\in \partial_{\alpha}(i)\setminus j }(1-\sigma_{\ell'\to i}^{\alpha})\right]\nonumber \\
	&&\times\prod_{\begin{subarray}{c}
        			\beta\\
			\beta\neq \alpha
      		\end{subarray}}\left[1-b_{i,\beta}\prod_{\ell\in \partial_{\beta}(i)}(1-\sigma_{\ell\to i}^{\beta})\right]
	\label{mes}
\eea
where ${\partial}_{\alpha}(i)$ indicates the nodes $\ell$ that are neighbors of node $i$ in layer $\alpha$ and the expression $\partial_{\alpha}(i)\setminus j $ indicates all the set of all the nodes belonging to $ \partial_{\alpha}(i) $ except node $j$.
Therefore the message {$\sigma_{i\to j}^{\alpha}$} is equal to one, if the node has not been damaged ($s_{i}=1$), if at least a message coming from a neighbor node $(\ell',\alpha)$ different from $(j,\alpha)$ is positive {(first factor in the square brackets)}, and if in all the layers $\beta$ where node $i$ is active ($b_{i,\beta}=1$), there is at least one neighbor $(\ell,\beta)$ of node $(i,\beta)$ sending a positive  message to node $(i,\beta)$ {(product over $\beta$)}.
Finally, the  indicator function that node $i$ active in layer $\alpha$, is in the MCGC is given by 
\bea
	S_{i\alpha}&=&s_{i} \left[1-\prod_{\ell'\in \partial_{\alpha}(i) }(1-\sigma_{\ell'\to i}^{\alpha})\right]\nonumber \\
	&&\times  \prod_{\begin{subarray}{c}
        			\beta\\
			\beta\neq \alpha
      		\end{subarray}}\left[1-b_{i,\beta}\prod_{\ell\in \partial_{\beta}(i)}(1-\sigma_{\ell\to i}^{\beta})\right] .
	\label{Smes}
\eea
This indicator function {  equals} one, if the node $(i,\alpha)$ has not been damaged ($s_{i}=1$), if at least a message coming from a neighbor node $(\ell',\alpha)$ of layer $\alpha$ is positive, and if in all the layers $\beta$ where node $i$ is active ($b_{i,\beta}=1$), there is at least one neighbor $(\ell,\beta)$ of node $(i,\beta)$ sending a positive  message to node $(i,\beta)$.
{In this section we only consider {\it uncorrelated activity networks} in order to allow for a theoretical treatment of the percolation properties of the multiplex networks with heterogeneous activity of the nodes.
Moreover we assume that the number of nodes $N_{\alpha}$ is large in every network $\alpha$ of the multiplex network.
Given that from Eq. $(\ref{p-factorizable})$ we have 
\bea
p_{i,\alpha}=\frac{B_i N_{\alpha}}{\avg{B}N}\leq 1,
\eea
we have 
\bea
1 \ll N_{\alpha} \leq \frac{\avg{B}N}{B_i},
\eea
where the condition $N_{\alpha}\gg 1 $ is required to study the percolation transition of the MCGC.
Since the value of the activity of the nodes $B_i\leq M$ we find the condition
\bea
\frac{\avg{B}N}{M} \gg 1.
\eea
This means that, for example, for finite $\avg{B}$ the number of layers {  is much smaller} than the total number of nodes.

In order to characterize how the size of the mutually connected giant component depends on the distribution of activities $P(B)$, here we consider the ensemble of multiplex networks with given activities of the nodes described in section \ref{ensemble}, where we assume additionally that the damage occur on a node $i$ with probability $1-p$, i.e. 
\bea
P(\{s_i\})=\prod_{i=1}^N p^{s_i}(1-p)^{1-s_i}.
\eea

Let us observe now that the probability that a node $i$ active in layer $\alpha$ has activity $B_i=B$ is given by 
\bea
P(B_i=B|b_{i,\alpha}=1)&=&\frac{P(B_i=B,b_{i,\alpha}=1)}{P(b_{i,\alpha}=1)}\nonumber \\
&=&P(B)\frac{BN_{\alpha}}{\avg{B}N}\left[\sum_B P(B) \frac{BN_{\alpha}}{\avg{B}N}\right]^{-1}\nonumber \\
&=&\frac{P(B) B}{\avg{B}},
\label{pbd}
\eea
where we have assumed that $p_{i,\alpha}$ is given by Eq. $(\ref{p-factorizable})$.
Assuming that all the layers have the same activity $N_{\alpha}=N \ \forall \beta$, and the same degree distribution $P(k)$,
by averaging the messages given by Eq.~$(\ref{mes})$ over the ensemble of multiplex networks with given activities of the nodes,  we  get the probability $\sigma^{\alpha}=\Avg{\sigma^{\alpha}_{i\to j}}$ that by following a link in layer $\alpha$ we reach a node in the mutually connected component}, obtaining $\sigma^{\alpha}=\sigma$ and  \bea
	\hspace{-7mm}\sigma&=&p\sum_B \frac{B P(B)}{\avg{B}}  [1-G_1(1-\sigma)] \left[1- G_0(1-\sigma)\right]^{B-1},
	\label{sa2}
\eea 
where the generating functions $G_0(x)$ and $G_1(x)$ are given by 
\bea
	G_0(x)&=&\sum_k P(k)x^k,\nonumber \\
	G_1(x)&=&\sum_k \frac{k}{\avg{k}}P(k)x^{k-1}.
\eea

Which can be derived as in the following. First we define $\sigma^{\alpha}$ as
\bea
\sigma=\sigma^{\alpha}&=&\Avg{\sigma_{i\to j}^{\alpha}}=\sum_B  P(B_i=B | b_{i,\alpha}=1)\left.\Avg{ \sigma_{i \to j}^{\alpha}}\right|_{B_i=B, b_{i,\alpha}=1}\nonumber
\label{snp1}
\eea
where $\Avg{\ldots}$ indicates the average over the network, in the large network limit. Using Eq. $(\ref{mes})$ we have
\bea
\sigma&=&p[1-G_1(1-\sigma)]  \sum_B P(B_i=B|b_{i,\alpha}=1)	\nonumber \\
&&\times \left.\Avg{\prod_{\beta \neq \alpha}\left[1-b_{i,\beta}\prod_{\ell\in \partial_{\beta}(i)}(1-\sigma_{\ell\to i}^{\beta})\right]}\right|_{B_i=B, b_{i,\alpha}=1} \\
&=& p[1-G_1(1-\sigma)] \sum_B \frac{P(B) B}{\avg{B}} [1-G_0(1-\sigma)]^{B-1}. \nonumber
\eea 
getting Eq. $(\ref{sa2})$ in the tree-like local approximation.

Using a similar derivation, under similar assumptions, the fraction $S^{\alpha}=\Avg{S_{i\alpha}}=S$ of the $N$ nodes that are in the mutually connected component in layer $\alpha$ is given by 
\bea
\hspace{-7mm}S&=&p\sum_B \frac{B P(B)}{\avg{B}}  [1-G_0(1-\sigma)]\left[1-G_0(1-\sigma)\right]^{B-1}.
\eea

These equations can be studied in detail as a function of the degree distribution in each layer, and the node activities distribution $P(B)$.
In particular, Eq.~$(\ref{sa2})$ can be expressed as,
\bea
	\sigma=p [1-G_1(1-\sigma)] K\left(1-G_0(1-\sigma)\right),
\label{uno}
\eea
where we have indicated by  $K(x)$ the following generating function
\bea
K(x)&=&\sum_B \frac{B}{\avg{B}} P(B)x^{B-1}.
\label{K}
\eea
Similarly $S^{\alpha}=S$ can be written as 
\bea
S=p [1-G_0(1-\sigma)]  K\left({1-G_0(1-\sigma)}\right).
\label{eq:S_er}
\eea

Let us consider now Poisson  networks identical on each layer: $P(k) = e^{-c} c^k/k!$ .Then we have $G_0(z) = G_1(z) = e^{-c(1-z)}$ and  the equation for $\sigma$ becomes $h(x)=0$ where
\bea
	h(x) = x-\tilde{p} (1-e^{-x})  K(1-e^{-x}),
	\label{eq:hx}
\eea
and $x=c\sigma$, $\tilde{p}=cp$.
A discontinuous transition occurs only if $h(x)$ has  maximum, so we need to impose the condition  $h(x)=h'(x)=0$, where
\bea
	h^{\prime}(x)& =& 1-\tilde{p}e^{-x} \left[ K\left({1-e^{-x}}\right) +(1-e^{-x}) K^{\prime}\left({1-e^{-x}}\right) \right],\nonumber
	\label{eq:hprimex}
\eea
where the function $K^{\prime}(x)$ is given by 
\bea
K^{\prime}(x)&=&\sum_B \frac{B(B-1)}{\avg{B}}P(B)x^{B-2}.
\label{J}
\eea
Using this equation to isolate $\tilde{p}$, and substituting into $h(x)=0$, we get for $x\neq 0$
\bea
	\frac{K^{\prime}\left({1-e^{-x}}\right)}{K\left({1-e^{-x}}\right)}&=&\frac{e^x}{x}-\frac{1}{1-e^{-x}}.
	\label{eq:KpK}
\eea

As a general remark, we expect in this model that the nature of the percolation phase transition will be dependent on the fraction of replica nodes with activity $B=1$. In fact these nodes do not have any interdependency on replica nodes in other layers. Therefore a high density of nodes with activity $B=1$ favors the continuous phase transition. In the extreme case in which  all the nodes have activity $B=1$ Eq. (\ref{eq:hx}) is in fact reduced to the equation determining the giant component of a single network. In fact, the generating function $K(x)$ defined in the last equation of $(\ref{K})$ is equal to one if all the nodes have activity $B=1$.  When the not all the nodes have activity $B=1$, the non trivial functional form of  $K(x)$ is determined by the nodes that have activity $B>1$.
Therefore we expect that the density of such nodes is sufficiently high the transition is continuous while otherwise,  we expect a discontinuous phase transition. This expected phenomenon can be related with a similar effect generated by the partial interdependence in multiplex networks where all the nodes are active in any layers \cite{Havlin2,Son}. 

In the following we provide evidence for this observations by considering two different distributions of activities $P(B)$: a scale-free activity distribution and a Poisson activity distribution.
\begin{figure}
	\includegraphics[width=0.95\columnwidth]{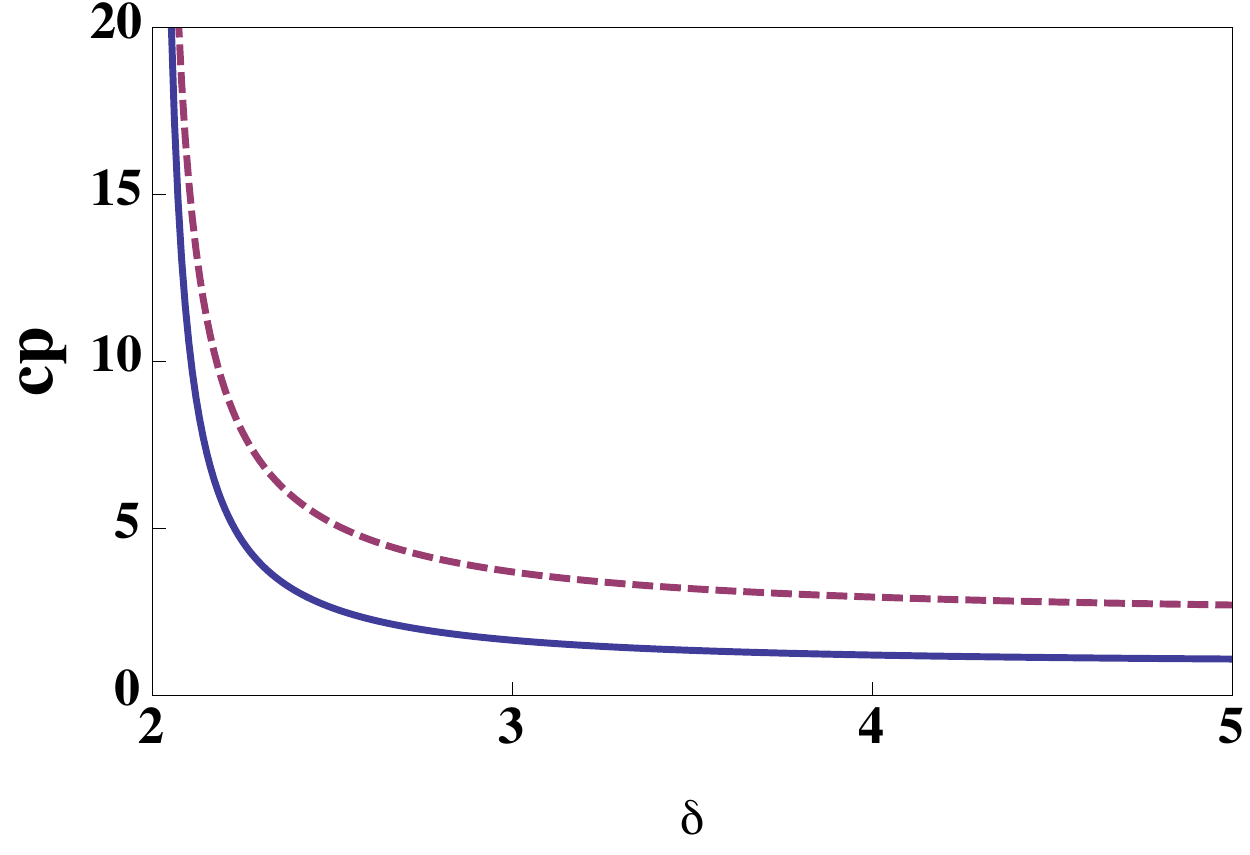}
	\caption{Phase diagram of a multiplex network formed by Poisson layers in the case of a power law activity distribution with exponent $\delta$. The solid line indicates the continuous phase transition obtained for $B_{min}=1$, and the dashed line indicates the discontinuous phase transition obtained for $B_{min}=2$. }
	\label{fig:sf-symm-ph-diag}
\end{figure}

\begin{figure}
	\includegraphics[width=0.95\columnwidth]{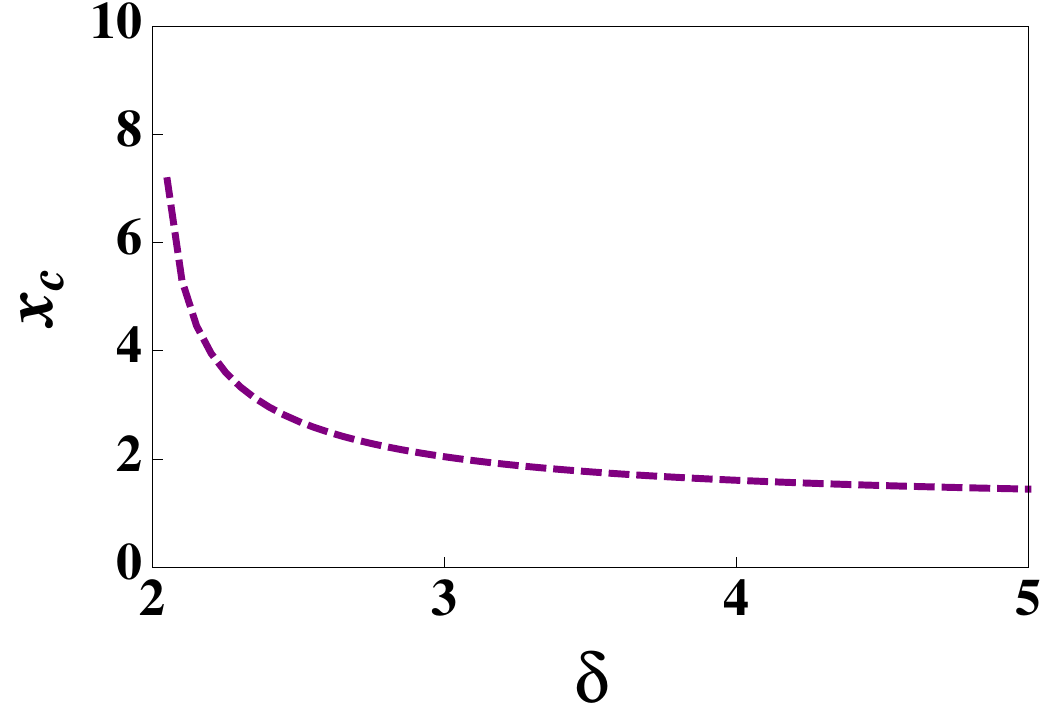}
	\caption{ Jump in $x_c=c\sigma_c=cS_c$, where $S_c$ is the size of  the MCGC at the discontinuous transition for a multiplex network formed by Poisson layers with mean degree $c$,  in the case of a power law activity distribution with exponent $\delta$ and  $B_{min}=2$. }
	\label{fig:sf-xc}
\end{figure}

\subsection{Scale-free $P(B)$ distribution}
Let us consider a power law activity distribution $P(B) = B^{-\delta}/\sum_{B=B_{min}}^{\infty}B^{-\delta}$ with $B\geq B_{min}$.
In this case, the function $h(x)$ is given by Eq. (\ref{eq:hx}) where 
\bea
	K(x) = \frac{\sum_{B=B_{min}}^{\infty}B^{1-\delta}x^{B-1}}{\sum_{B=B_{min}}^{\infty}B^{1-\delta}}
\eea

The position of the discontinuous transition can be calculated by solving  $h(x^{\star})=h^{\prime}(x^{\star})=0$, with $x^{\star}>0$.
The discontinuity of the transition is due to the finite value of the solution at $x^{\star}>0$, as one can see that $S>0$ from Eq.~(\ref{eq:S_er}), where  $G_0(z)$ is a simple exponential.
The position of the continuous transition can be instead calculated by solving $h(0)=h^{\prime}(0)=0$.
While the tricritical point, if it exist, can be found by setting $h(0)=h^{\prime}(0)=h^{\prime\prime}(0)=0$.
(Note, that given the functional form of $h(x)$, given by Eq. $(\ref{eq:hx})$, the condition $h(0)=0$ is always satisfied.) 

Fig.~\ref{fig:sf-symm-ph-diag} illustrates the phase diagram  calculated for $B_{min}=2$ and $B_{min}=1$. 
For $B_{min}=1$ we have  a continuous phase transition for $\delta>2$.
For $B_{min}=2$ we find a discontinuous phase transition line across all the values of $\delta>2$.
The jump in the size of the MCGC $x_c=c\sigma_c=cS_c$ at the discontinuous transition for $B_{min}=2$ (shown in Fig.~\ref{fig:sf-xc}) diverges as $\delta$ approaches the value 2 (while at the continuous transition there is no jump).
Both transition lines for $B_{min}=1$ and $B_{min}=2$ diverge as $\delta$ approaches the value $2$, indicating that  as $\delta$ decreases the  multiplex network becomes more fragile.
As $\delta$ decreases, the fraction of nodes with large activity (activity hubs) becomes important. These nodes, to be in the mutually connected giant component they must be connected in each layer, and if they are damaged they affect the connectivity of multiple layers. Therefore they are at the same time more fragile than nodes with smaller activity and are able to  affect the multiplex connectivity across more layers.
As a result, as $\delta$ decreases,  random damage affects more and more layers and the multiplex network becomes less robust.

\subsection{Poisson $P(B)$ distribution}
For activities that are Poisson distributed $P(B)=\mu^Be^{-\mu}/B!$,  we have 
\bea
	h(x) = x-\tilde{p} (1-e^{-x}) \exp\left[ -\mu ({e^{-x}} ) \right].
	\label{eq:poisson-activity-h}
\eea

\begin{figure}
	\includegraphics[width=0.9\columnwidth]{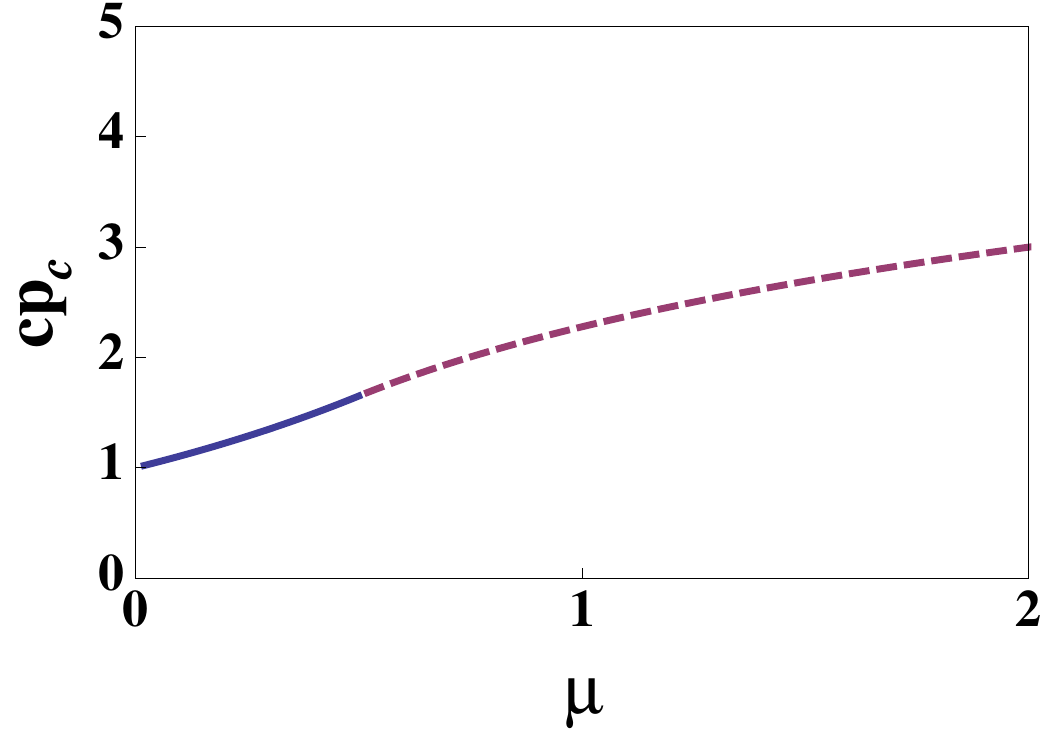}
	\caption{Phase diagram of a multiplex network formed by Poisson layers with identical mean degree $c$ and with a Poisson activity distribution with mean $\mu$. The blue solid line is a line of continuous phase transitions between a percolating and a non-percolating phase, the red dashed line represents  instead the line of discontinuous phase transitions.}
	\label{fig:poisson}
\end{figure}

\begin{figure}
	\includegraphics[width=0.9\columnwidth]{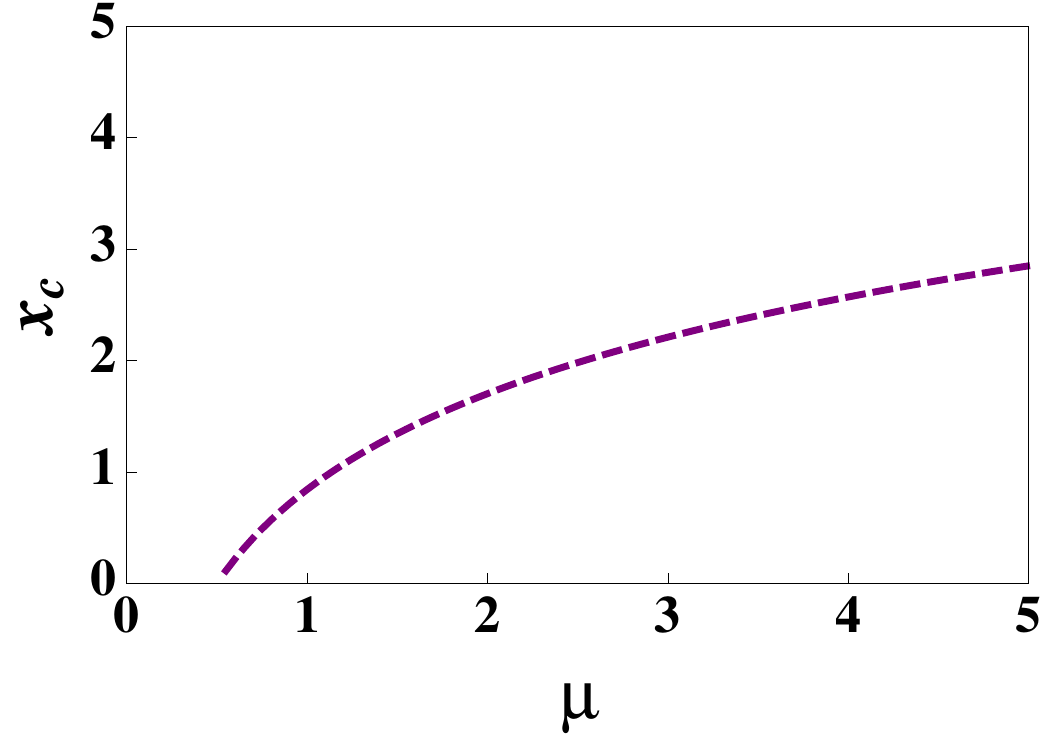}
	\caption{ The jump in $x_c=c S_c$  where $S_c$ is  the size of the mutually connected giant component at the discontinuous transition point, is shown for a  multiplex network formed by Poisson layers with identical mean degree $c$ and with a Poisson activity distribution with mean $\mu$.}
	\label{fig:poisson_xc}
\end{figure}

By setting $h(x^{\star})=h^{\prime}(x^{\star})=0$, we can find a line of discontinuous transitions that ends at a critical point defined by $h(0)=h^{\prime}(0)=h^{\prime\prime}(0)=0$ characterized by 
\bea
	\tilde{p}=\tilde{p}_T = \sqrt{e}\simeq 1.648, \qquad \mu=\mu_T = \frac{1}{2}=0.5
	\label{poisson-tricritical}
\eea
In the multiplex network, when $\mu<\mu_T$ the percolation transition is continuous. Instead for $\mu>\mu_T$ the percolation transition is discontinuous.
In Figure $\ref{fig:poisson}$ we show the phase diagram.
At $\mu=0$, the continuous phase transition is at $\tilde{p}=pc=1$ as expected for independent layers. As the level of interdependence becomes increasingly significant, the continuous transition becomes discontinuous (for $\mu>\mu_T$). 
This can also be seen from the size of the discontinuous jump $x_c=c\sigma_c=cS_c$, that approaches zero at the tricritical point  $\mu_T$ and is shown in Figure $\ref{fig:poisson_xc}$. 
%It is interesting to compare (Fig.~\ref{fig:poi-sf-comparison-ph-diag}) the resilience of two multiplex networks with different activity distributions (Poisson and power-law), but with the same mean $\mu=\zeta(\delta-1)/\zeta(\delta)$.
%At high $\delta$ ($\mu\to 1$), the power-law activity distribution is more robust than the Poisson one, as the hubs keep the multiplex networks connected and are not efficiently targeted by random damage.
%As $\delta\to 2$ ($\mu\to \infty$), however, the transition line of the power-law activity distribution diverges more rapidly than the Poisson one, as the layers become more and more interdependent.
%%
%\begin{figure}
%	\includegraphics[width=0.95\columnwidth]{comp.pdf}
%	\caption{Phase diagram of multiplex  networks formed by Poisson layers. Comparison between the case of Poisson (red, dashed line) and power-law (blue, solid line) activity distribution with the same mean $\mu=\zeta(\delta-1)/\zeta(\delta)$.}
%	\label{fig:poi-sf-comparison-ph-diag}
%\end{figure}
%

\section{Correlations between the activities and the degrees of the nodes}

In this section, we consider the role of correlations between the activities of the nodes and their degrees in the layer in which they are active.
The higher is the activity of a node, the more fragile is the node, but the higher is the degree of the node within the layer the more robust it is.
Therefore, here we want to characterize the effect that the correlations between the activity and the degrees of  each node have on the robustness properties of the entire multiplex network.
We start from the message passing Eqs.~$(\ref{mes})$ and $(\ref{Smes})$. We assign to each node $i$ an activity $B_i$ from a $P(B)$ distribution.
Then, we assume that the matrix of activities is given by ${\bf b}$ with probability $P({\bf b})$
given by Eq.~$(\ref{Pben})$ with $p_{i,\alpha}$ given by Eq.~$(\ref{p-factorizable})$.
Finally, we need to extend Eq.~(\ref{Pkb}) for the degree sequence to the correlated case.
As node degrees on each layer are correlated with node activities,  the probability $P(\{\bf k\}|{\bf b},{\bf B})$  that a multiplex network has degree sequences $\{\bf k\}_{i,\alpha}$, given the activity matrix ${\bf b}$ and the activity sequence $\{B_i\}_{i=1,2\ldots,N}$, is 
\bea
	P(\{\bf k\}|{\bf b},{\bf B})=\prod_{i,\alpha}\left[P_{B_i}(k_i^{\alpha}){b_{i,\alpha}}+\delta(k_i^{\alpha},0)(1-b_{i,\alpha})\right].\nonumber\\
\eea 
In particular, we assume for simplicity that
\bea
	P_{B_i}(k)=\frac{1}{k!}c(B)^k e^{-c(B)}
	\label{poisson-correlated}
\eea
and we take 
\bea
	c(B)=c_0 B^a
	\label{B-power-law}
\eea
where $c_0,a$ are two parameters determining the correlations between the degrees of the node and its activity.
The particular choice of  the functional form of $P_{B_i}(k)$ in Eqs.~$(\ref{poisson-correlated})$-$(\ref{B-power-law})$ is dictated by the intention to model positive correlations between the activity of the degree of the layers that have been observed in real dataset \cite{Vito}. Real multiplex network analysis are nevertheless not sufficient to suggest the exact form of the correlations observed. Therefore this  ensemble of correlated multiplex networks with heterogeneous activity of the nodes has been chosen in such way  describe positive correlations between activities and degree in single layers, while  keeping the model sufficiently simple to allow   a number of analytical calculations.
%\begin{figure}
%%	\includegraphics[width=0.95\columnwidth]{corr.pdf}
%	\caption{ Percolation threshold $p_c$ of a multiplex network with correlation between the activity and the degree of the nodes in single layers. The activity distribution is a power law with exponent $\delta=2.5$. As the correlation between the activity and the degree in each layer increases, {\ie} as the parameter $a$ increases, the network becomes more robust and the percolation threshold decreases. }
%	\label{fig:corr}
%\end{figure}
%\begin{figure}
%	\includegraphics[width=0.95\columnwidth]{adelta-nonzero-cp-dav.pdf}
%	\caption{ Line of tricritical points for  a multiplex network in the case of a  activity distribution $P(B)\propto B^{-\delta}$ for $B\in[1,100]$  and degree in within each layer correlated with the activity of the node according to Eqs.$(\ref{poisson-correlated})-(\ref{B-power-law})$. }
%	\label{fig:tricritical}
%\end{figure}

\begin{figure}
	\includegraphics[width=0.95\columnwidth]{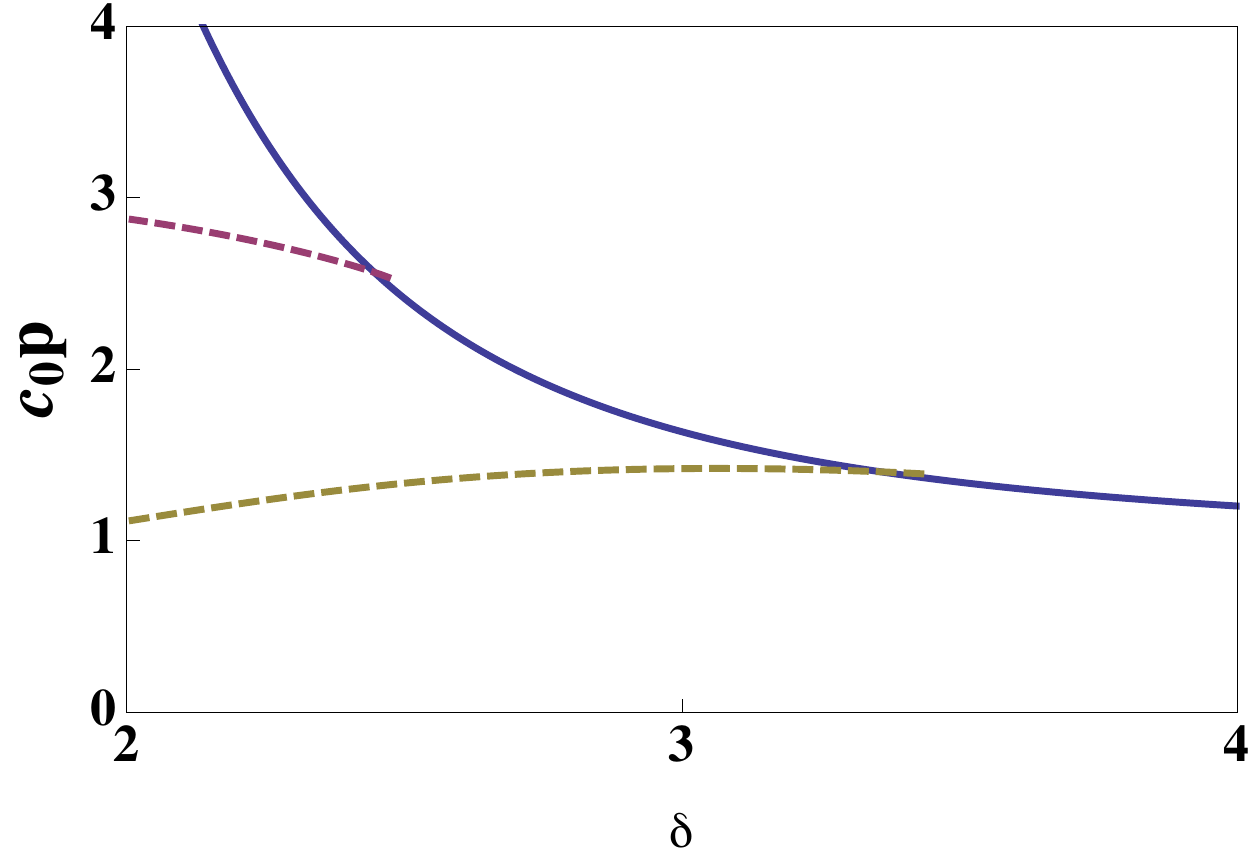}
	\caption{Phase diagram of a multiplex network in the case of an  activity distribution $P(B)\propto B^{-\delta}$ for $B\in[1,100]$  and degree  within each layer correlated with the activity of the node according to Eqs.~$(\ref{poisson-correlated})-(\ref{B-power-law})$. From top to bottom,  the dashed lines indicate discontinuous  phase transitions for $a=0.2$ and for $a=0.6$.  Such lines delimit a line of continuous transitions that does not depend on $a$. The discontinuous lines end into a critical point.
	The plot shows that  the stronger are the correlations between the activity of the nodes and their degree within  each layer, the larger is the percolating phase, but at the same time the transition becomes discontinuous. }
	\label{fig:corrB1}
\end{figure}

\begin{figure}
	\includegraphics[width=0.95\columnwidth]{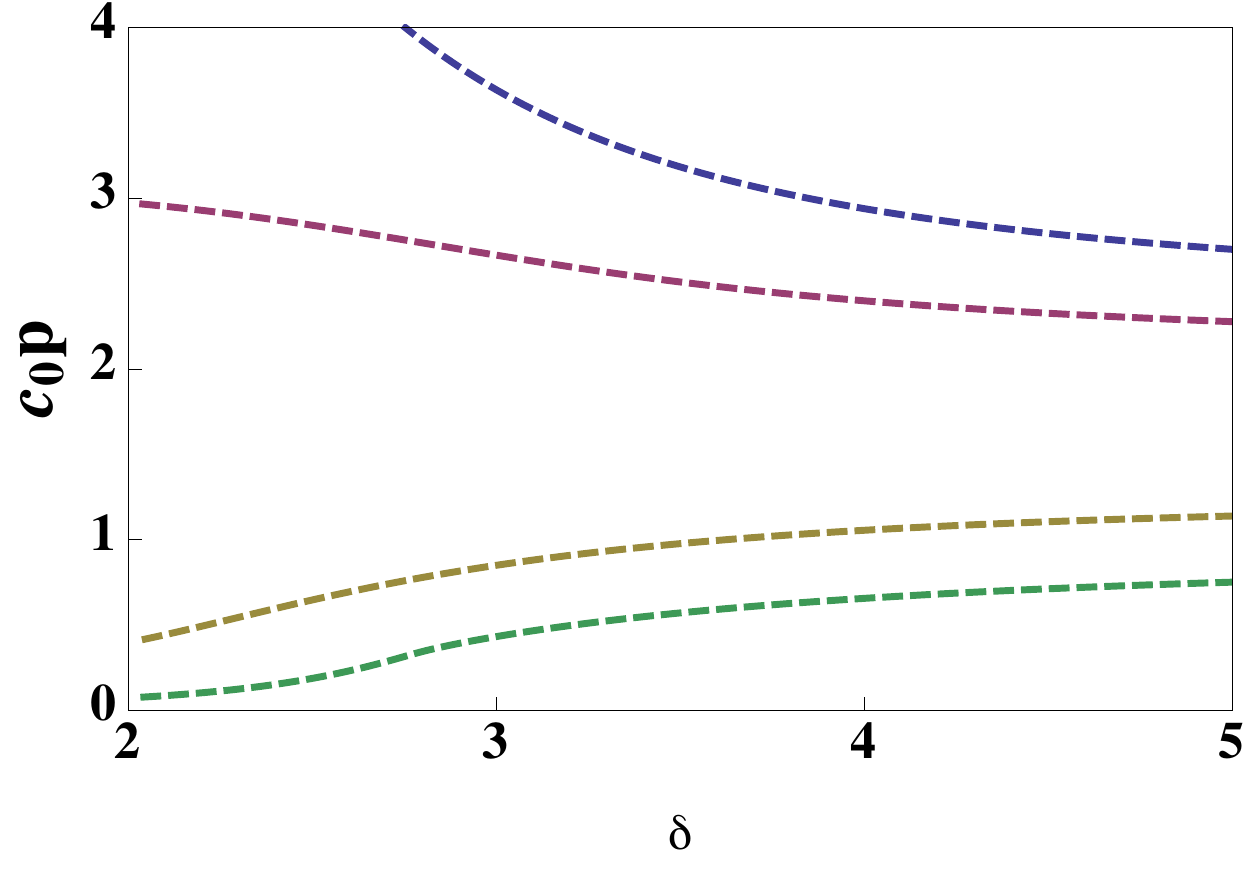}
	\caption{Phase diagram of a multiplex network in the case of an  activity distribution $P(B)\propto B^{-\delta}$ for $B\in[2,100]$  and degree within each layer correlated with the activity of the node according to Eqs.~$(\ref{poisson-correlated})-(\ref{B-power-law})$. From top to bottom, the  dashed lines  indicate discontinuous phase transitions for $a=0$, $a=0.2$, $a=1$ and $a=1.5$, respectively. Differently from the previous case, we do not have continuous transitions.
	 }
	\label{fig:corrB2}
\end{figure}

\begin{figure}
	\includegraphics[width=0.95\columnwidth]{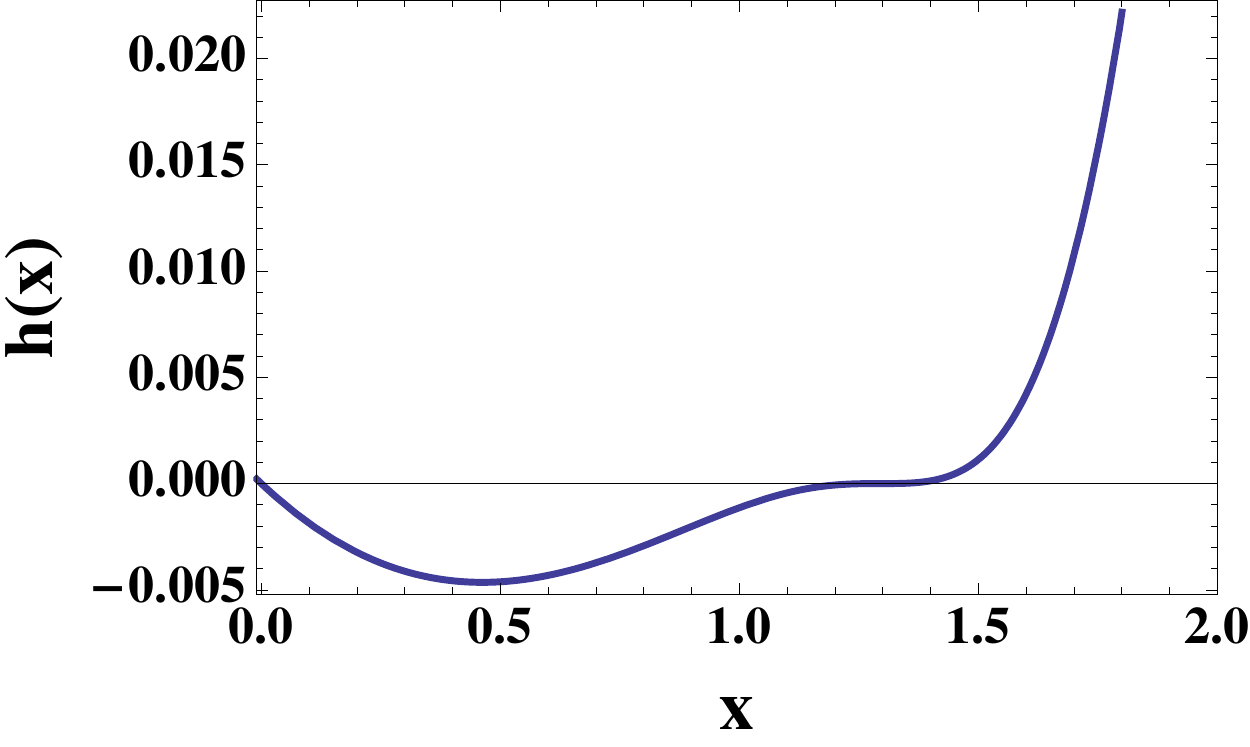}
	\caption{Solutions of Eq.~(\ref{hx}) at the critical point for $B_{min}=1$ and $a=0.2$.}
	\label{fig:solutions-hx}
\end{figure}

\begin{figure}
	\includegraphics[width=0.95\columnwidth]{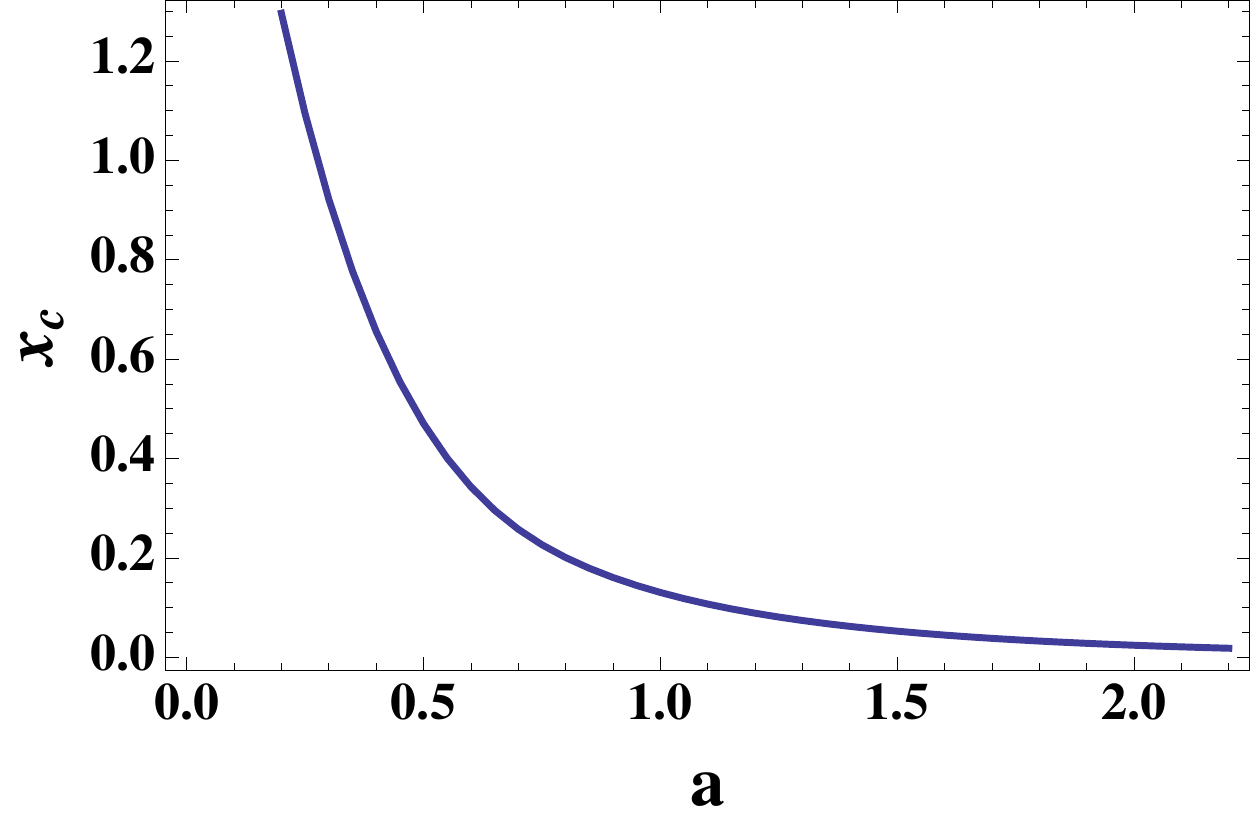}
	\caption{ Position of critical points for  a multiplex network in the case of an  activity distribution $P(B)\propto B^{-\delta}$ for $B\in[1,100]$  and degree on each layer correlated with the activity of the node according to Eqs.$(\ref{poisson-correlated})-(\ref{B-power-law})$. }
	\label{fig:critical-values}
\end{figure}

Averaging the message passing equation over this ensemble and assuming $N_{\alpha}=N$, $\forall \alpha$, and $B\ll M$, we have only one average message determined by the equation
\bea
	S=\sigma=p\sum_{B}\frac{B}{\avg{B}}P(B)\left(1-e^{-c(B)\sigma}\right)^{B-1}\left(1-e^{-c(B)\sigma}\right).
\eea 
Setting $\tilde{p}=c_0p$, the equation for $x=c_0\sigma$ reads
\bea
h(x)=0
\eea
with
\bea
	h(x)=x-\tilde{p}\sum_{B}\frac{B}{\avg{B}}P(B)\left(1-e^{-B^ax}\right)^{B-1}\left(1-e^{-B^a x}\right).\nonumber
	\label{hx}
\eea
This equation can be studied as a function of $a$, $\tilde{p}$ and the parameters determining the $P(B)$ distribution.
In order to find the discontinuous phase transition, we set $h(x^{\star})=0$, $h^{\prime}(x^{\star})=0$.
The line of these discontinuous phase transition eventually stops at a  critical point $x_c$, that can be calculated by setting 
\bea
h(x_c)=h^{\prime}(x_c)=h^{\prime\prime}(x_c)=0.
\label{xceq}
\eea
%and obtain the following system of equations, respectively
%%\begin{widetext}
%\bea
%	\frac{1}{\tilde{p}} &=& \sum_{B}\frac{B}{\avg{B}}P(B) B^{a}e^{-B^a x} e^{-(B-1)e^{-B^ax}}\times \nonumber\\
%	&&\times\left[B-(B-1)e^{-B^a x}\right]
%\eea
%\bea
%	0 &=& \sum_{B}\frac{B}{\avg{B}}P(B) B^{2a} e^{-(B-1)e^{-B^ax}}  \times\nonumber \\
%	&& \left[ (1-2B+B^2)e^{-3B^ax} \right. \nonumber \\
%	  &&  \left.+  (2-B-B^2) e^{-2B^a x}+ Be^{-B^a x} \right].
%\eea
%%\end{widetext}
The continuous phase transition can be found, instead, by imposing $h(0)=h^{\prime}(0)=0$.
%The line of these second order critical points end at the tricritical points can be calculated by imposing \bea
%h(0)=h^{\prime}(0)=h^{\prime\prime}(0)=0.\label{tricritical}\eea
%This corresponds to $a=a_T$, where $a_T$ must satisfy the  equation 
%\bea
%\sum_{B}\frac{BP(B)}{\avg{B}}e^{-(B-1)}B^{2a_T}(2B-3)=0.
%\eea
%From this equation it transpires that if the minimal activity is $B_{min}\geq 2$, the transition is always discontinuous and we do not have any tricritical point.

Let us now consider a power-law distribution of activities $P(B)\propto B^{-\delta}$ for $B\in[B_{min},100]$ and correlated degrees in every layer according to the degree distribution given by Eq.~(\ref{poisson-correlated}).
In Figure $\ref{fig:corrB1}$ and Figure $\ref{fig:corrB2}$ we show  the percolation transitions as a function of the parameter $a$ for  $B_{min}=1$ and $B_{min}=2$, respectively.
As a general remark, for $B_{min}=2$, the transition appears to be always discontinuous, while for $B_{min}=1$ we have both continuous and discontinuous transitions, with a region of coexistence between two percolating phases, and a critical point at the end of the line of discontinuous transitions.

First, let us observe the case of $B_{min}=1$ (Fig.~\ref{fig:corrB1}).
We do not plot the line $a=0$ as it reduces to the non-correlated case shown in Fig.~\ref{fig:sf-symm-ph-diag}, where the phase transition is always continuous.
For $a>0$, instead, it emerges a line of discontinuous phase transitions ending in a critical point, which is determined by the Eqs. $(\ref{xceq})$.
The line of continuous phase transitions encounters the line of discontinuous transitions at a critical end point determined by the simultaneous occurrence of two minima of function $h(x)$ (Eq.~\ref{hx}).
Fig.~\ref{fig:solutions-hx} shows the solution of equation (\ref{hx}) at the  critical point.
This phase diagram, therefore, is characterized by the presence of two percolating phases, separated by a short line of discontinuous transitions (Fig.~\ref{fig:corrB1}).
This coexistence region shifts towards larger values of $\delta$ as $a$ increases, but the critical point never joins the continuous line at a tricritical point at finite $a$.
Figure $\ref{fig:critical-values}$ shows that $x_c\to 0$, without joining the continuous solution at $x=0$, for $a\to\infty$.

In the case $B_{min}=2$ (Figure $\ref{fig:corrB2}$), we observe a discontinuous phase transition for all values of $a$.
This qualitative difference is due to the peculiar role of the nodes which are active on a single layer ($B=1$). In our model, this type of nodes do not need support from the other layers (as they appear only in one of them), and therefore they drive a classical continuous percolation transition.
In both cases, as $a$ increases, the percolating threshold becomes smaller, as the intensity of the correlations between the activities of the nodes and the degree of the nodes within each layer increases. Therefore the multiplex networks is more robust because it can still have a mutually connected component even if the damage is significant.
On the other hand, though, the phase transition becomes discontinuous, and therefore the collapse becomes more unpredictable.

%}

\section{Conclusions}

In this paper, we have characterized the robustness properties of multiplex networks in a new model that encapsulates heterogeneous activities of the nodes, {\ie} the possibility that each node is present only on a small fraction of layers  in a multiplex, as seen in real world cases \cite{Vito}.
In this model, we employ a notion of mutual percolation where nodes must belong to a mutually connected component only on the layers where they are active and develop an analytical approach to calculate the size of the mutually connected giant component as a function of node damage and other parameters.
We show that multiplex networks with very broad activity distributions are more fragile than networks with more homogeneous distribution of activities.
%, when multiplex with the same mean activity are compared.
%Moreover, we show that correlations between the activities of the nodes and their degrees in single layers enhance the robustness of multiplex networks to the extent the percolation threshold corresponds to higher fraction of initially damaged nodes.
%This provides an example on how correlations can typically reduce the fragility of multiplex networks and contribute to reduce the dramatic effects of global disruptions.

We also investigate the role of correlations between the activities of the nodes and their degrees.
We show that these correlations generally improve the stability of the percolating phase,and the multiplex network has a smaller percolation threshold, so the multiplex network becomes more robust.
However, correlations also change the order of the phase transition, that becomes discontinuous.
This provides an example on how correlations can typically reduce the fragility of multiplex networks, but at the same time they can make the system more unpredictable, as the transition becomes discontinuous.

%==============================================================
\section{Acknowledgments}
We acknowledge useful discussions with  Vincenzo Nicosia and Vito Latora.
This work has been partially funded by: Science Foundation Ireland, grants 14/IF/2461 and 11/PI/1026; the FET-Proactive project PLEXMATH (FP7-ICT-2011-8; grant 317614).

\end{document}